# Proving weak electronic interaction between molecules and substrate: a study of pentacene monolayer on graphite

Yuri Hasegawa[1, +], Takuma Yamaguchi[2], Matthias Meissner[1], Takahiro Ueba[1], Fabio Bossolotti[1], Shin-ichiro Ideta[3, ++], Kiyohisa Tanaka[2, 3], Susumu Yanagisawa[4], Satoshi Kera[1, 2, 3 *]

[1] Department of Photo-Molecular Science, Institute for Molecular Science, Okazaki 444-8585, Japan
[2] SOKENDAI (The Graduate University for Advanced Studies), Okazaki 444-8585, Japan
Japan
[3] UVSOR Synchrotron Facility, Institute for Molecular Science, Okazaki 444-8585, Japan
[4] Department of Physics and Earth Sciences, Faculty of Science, University of the Ryukyus, University of the Ryukyus, Nishihara, Okinawa 903-0213, Japan

[+] Present address: Department of Physical Sciences, Ritsumeikan University, Kusatsu, 525-8577, Japan
[++] Present address: Hiroshima Synchrotron Radiation Center, Hiroshima University, Hiroshima, 739-8511, Japan

[*]Corresponding author: kera@ims.ac.jp

**Abstract**

The impact of van der Waals interaction on the electronic structure between a pentacene monolayer and a graphite surface was investigated. Upon cooling the monolayer, newly formed dispersive bands, showing the constant final state nature overlapping with the non-dispersive, discrete molecular orbital state, is observed by low-energy angle-resolved photoelectron spectroscopy. The dispersive band consists of positive and negative intensities depending on the final state energy, indicating Fano resonance involving a discrete molecular state that couples a continuum state upon photoionization. A wave-function overlap is demonstrated according to their larger spread in unoccupied states even at the weakly bounded interface by Fano profile analysis.

**Introduction**

The interaction between an organic molecule and an inorganic surface is important to understand the mechanism of forming interfacial electronic states. Among the systems of adsorption-induced modifications of the electronic state, a metal substrate-induced modification of molecular orbitals has been demonstrated intensively, while the impact of weak interaction on the electronic state has not yet been investigated because of fewer interests by a simple approximation of van der Waals (vdW) interaction. To unveil a rich assort of peculiar properties and unique functionalities given by molecular materials, however, it is necessary to understand rigorously the true characteristic of electronic states at weakly coupled electronic systems.

We have shown that weak interfacial interaction by vdW force affects the electronic state [1–3]. As an example, a pentacene (PEN) molecule on a graphite substrate is one of the famous systems stabilized by weak physisorption. Yamane *et. al.* studied the electronic structure of the PEN monolayer (ML) on graphite by angle-resolved ultraviolet photoelectron spectroscopy (ARUPS). There are differences in the spectral fine features of the highest occupied molecular orbital (HOMO) state to those of an isolated gaseous molecule, demonstrating the charge reorganization energy becomes large for the solid state [4]. Paramonov *et. al.* simulated the photoemission spectrum considering interfacial interaction and obtained a reasonably good agreement between the simulation and the experiment, suggesting the importance of weak interfacial interaction on delocalization of the discrete state as well as hole-vibration coupling [5]. These results demonstrate that a charge is highly localized in a molecule at the monolayer, where the theory shows a weak 11-meV dispersion for the HOMO band. Other non-trivial findings from them are that i) photoelectron-emission angle dependence on the vibronic progression is observed against Franck-Condon (FC) principle [2,4] and ii) a large density of states (DOS) is detected around the normal emission geometry in the ARUPS which is absent in a simulation of photoemission orbital tomography of an isolated molecule [6,7]. All these results inform us that deep insight into the fine features of ARUPS would be necessary to fully understand the electronic structure of weakly interacting molecular systems beyond the simple approximation.

Against a tiny change in the occupied valence states of the physisorption system, the unoccupied states would be expected to give many impacts on the spectral features due to a larger spread and overlapping of the electron cloud. Experimental access to band mapping of the unoccupied state is limited, however, it has been attempted by two-photon photoemission spectroscopy and vdW-corrected density functional theory (DFT) calculation that the image potential states of graphite interacting with the diffuse unoccupied state of naphthalene [8] and PbPc [9] play an important role to form the interfacial state.

In general, ARUPS is known to detect the band structure of the occupied state. On the other hand, the unoccupied state is accessible in some specific cases by so-called angle-resolved secondary-photoelectron spectroscopy (ARSEE) [10,11], which corresponds to the result of very low-energy electron diffraction (VLEED) experiment [12]. In ARUPS measurement, when the secondary electrons couple with the conduction band (CB) above the vacuum level ($E_v$), the electronic structure of the CB can appear to overlap with the valence band (VB) as the final state contribution showing a constant kinetic energy phenomenon. A constant final state spectrum (CFS) can be judged by scanning the photon energy to observe the feature moving with the excitation energy apart from a VB.

Fano resonance [13] is a universal phenomenon observed in many subjects. When a discrete state interferes with a broad continuum state, characteristic asymmetric feature such as Fano profile is found in the spectrum as a result of quantum interference [14]. In a photoelectron emission, several examples are reported for core excitation with the Auger decay process [15–18] and valence excitation of a rare gas superstructure on the surface [19], however no results on the complicated molecular orbital-related systems so far.

In this study, we measured the PEN (ML) on graphite substrates by low-energy ARUPS (LE-ARUPS) using a synchrotron radiation source that can access the band structure in the region where the secondary electron emission is dominant. The HOMO features of LE-ARUPS for PEN/graphite are entirely different from that of a conventional ARUPS using a HeI or HeII light source. LE-ARUPS of the close-packed molecular layer reveals that the convex (hole-like) dispersive band, which has a characteristic of CFS, appears to overlap with the discrete HOMO of PEN with vibronic fine features, while the convex band is absent in the valence band of pure graphite both for the experiment and theory [12,20,21]. The CFS dispersive band consists of positive and negative intensities depending on the excitation energy, indicating Fano resonance involving a discrete molecular state that couples a continuum state [13]. The continuum state at the PEN/graphite interface could be originated from a newly formed conduction band at the weakly bound interface, hence demonstrating that weak interaction becomes more impact on the hybridization of the wave functions of the conduction band. We make a concept to discuss the impact of weak interaction on the electronic states by Fano profile analysis.

**Results and discussion**

First, the temperature-dependent structural transition of the PEN in the sub-monolayer regime was measured. Figs. 1 (a) and (b) show LEED patterns taken at 72 K and 170 K, respectively. At 170 K, a filled hexagon-like pattern is observed, indicating the arrangement of molecules aligning with a mean distance from adjacent molecules and densely packed molecules with relatively small grains. Here we call this arrangement a liquid-like phase at a higher temperature (HT). On the other hand, a six-fold symmetry is observed at 72 K, indicating an apparent structural transition to the crystalline phase at a lower temperature (LT). The transition temperature between the LT and HT phases is around 130 K. In the LT phase, two slightly different patterns are observed depending on the coverage. There are several PEN(ML)/graphite structures reported. Fig. 1 (b) shows the LEED pattern with a nominal coverage of 3 Å (structure A). The other LEED pattern for a nominal coverage of 4 Å (structure B) is shown in Fig. S1 together with the previous results (structures C and D) [5,22]. The simulated LEED pattern reproduces well by the lattice parameter evaluated to be $a = 16.1$ Å, $b = 16.1$ Å, and $\beta = 60°$ as shown in Fig. 1(c). The structure A is commensurate packing within the experimental accuracy. Upon cooling the ML below the transition temperature, the change of electronic state is observed by UPS measurement with an excitation energy of 21 eV (Fig. 1(d)). At the HT phase, an asymmetric feature with a main sharp HOMO peak denoted (00) appears at 1.20 eV with progressing vibronic satellite peaks, (01) and (02), which are separated by approximately 0.16 eV, where (0$m$) describes a FC transition from a ground state to an excited state, and $m$ is a vibrational excited level. These spectral features are known to be originated from the hole-vibration coupling upon photoionization. At the LT phase, the whole spectrum shifts towards higher binding energy, and each line shape becomes sharper. There is a small peak located at 1.2 eV marked by "*". This additional component can be the remained HT. The shift of the HOMO feature upon the cooling indicates the relaxation energy by forming the electrostatic potential by the molecular quadrupole effect [23]. The line shape is nearly identical to that of a gaseous, while the reorganization energy is slightly affected by the substrates and the surrounding molecules as demonstrated in the theory for the interface model [5]. The angular dependence of the vibronic coupling intensity suggests breaking the sudden approximation [2]. A Lorentzian-tail contribution may give a lifetime of a photogenerated hole of the system [24]. The calculated band structures are almost identical for those structures shown in Fig. S2.

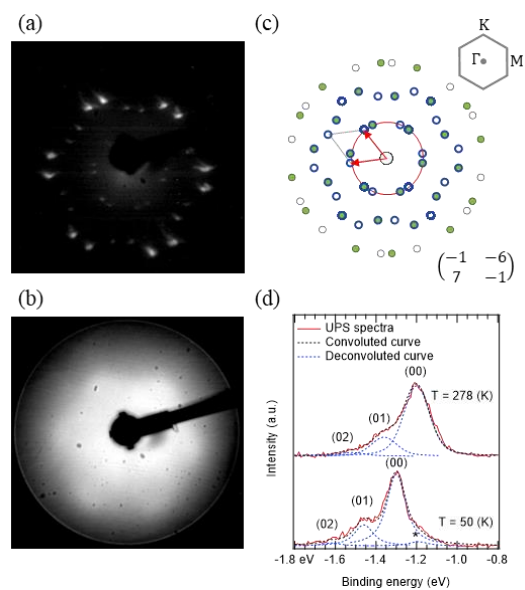

Fig. 1. LEED pattern of PEN/SCG taken at (a) 72 K (LT) and (b) 170 K (HT) with an electron beam energy of 18 eV. (c) Simulated diffraction pattern depicted considering rotational and mirror domains. The blue circles correspond to the diffraction spots observed experimentally. The surface brilluian zone of the graphite is shown. (d) ARPES spectra of PEN/HOPG corrected at 50 K (LT) and 278 K (HT) around K point where the photoemission intensity gives a maximum. The spectrum at 50 K (278 K) is reproduced by the Voigt function where 48 meV (119 meV) of Gaussian components and 91 meV (54 meV) of Lorentzian components, respectively.

Next, the photon energy dependence of the intensity map of energy and momentum (E-$k$) relation of PEN/HOPG and PEN/SCG were measured using LE-ARUPS for the LT and HT phases [1]. The photon energy dependence of the HOMO feature at the normal emission geometry shows unexpected features as for a physisorbed system. Figs. 2 (a) and (b) show the excitation energy dependence of the HOMO features of PEN/HOPG, and the intensity is normalized to the HOMO (01) peak. First, as described in the introduction a large DOS is observed around the normal-emission geometry, which is absent qualitatively for a system of planar π-conjugated molecules with flat-lying orientation in the ARUPS and also confirmed in the theory for a photoemission angular distribution of an isolated molecule [6,7]. In this study, PEN adsorbs almost flat-lying on the surface or inclining slightly the short axis to 10 ° as indicated by the optimized structure in the calculation based on the matrix obtained by the LEED pattern. Inclining PEN can show a faint photoemission intensity at the normal emission geometry but not strong as found in the experiments (see Fig. S3). Note that a small intensity observed around 0.95 eV at hν = 8.5 eV at 14 K exists as an unoccupied σ* band of pristine graphite [10]. Second, at both the LT and HT phases for hν = 8.5 eV, the FC factor of relative intensity between the HOMO (00) peak and the (01) or (02) satellite peak almost remain for a conventional HeI and HeII excitations, while that of other excitation energies varies and significantly different to the HOMO feature in the conventional ARUPS. The intensity abnormally of the vibronic progression of the HOMO at low energy excitation suggests that the FC principle is not satisfied in this system. On the other hand, the PEN/SCG system does not give a clear conclusion, because briefly the features of surface Umklapp scattering affect largely the ARUPS intensity distributions (see Fig. S4).

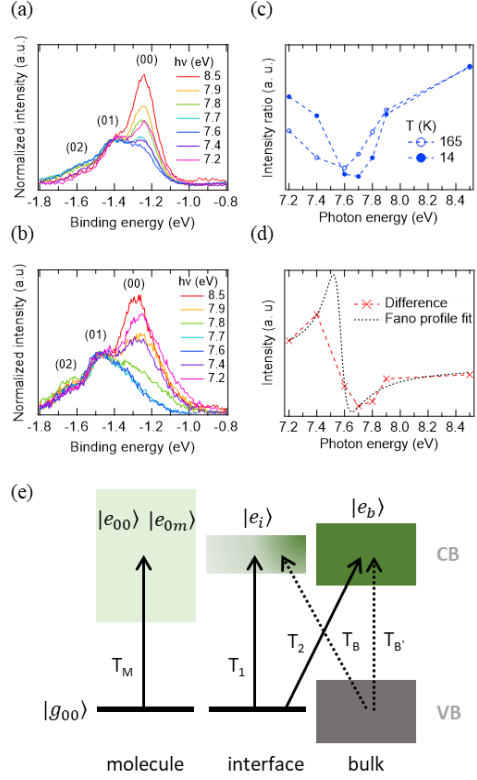

Fig. 2. Photon energy dependence of LE-ARUPS spectra of PEN/HOPG taken at (a) 14 K (LT) and (b) 165 K (HT) which is from a normal-emission geometry by integrated at k = 0.00 Å$^{-1}$ with $\Delta k$ = 0.04 Å$^{-1}$. The spectrum is shown after subtracting the background, and the intensity of each spectrum is normalized by the intensity at the HOMO (01) vibronic peak. (c) Photon energy dependence of the peak intensity of HOMO (00) in (a) and (b). (d) Photon energy dependence of the intensity obtained by dividing the spectra of T = 14 K by one of T = 165 K in (c). (e) Excitation schemes for the PEN/graphite in the LE photoelectron emission. Two possible resonant transitions are considered from a discrete-like HOMO to a broadened conduction band at the interface ($T_1$) and to the bulk states ($T_2$), which show Fano profile. The transition from the discrete-like HOMO to a free-electron state ($T_M$) gives a feature based on Franck-Condon principle with vibronic progressions. The transition from the bulk VB of the graphite to the bulk CB ($T_{B'}$) as well as to the interface state ($T_B$) could be neglected in the present results.

To compare the LT and HT phases, the relative intensity of (00) / (01) peak is plotted as a function of the photon energy (Fig. 2(c)). Here, the relative intensity of HOMO drops to around 7.7 eV in the LT phase. On the other hand, the tendency is slightly different in the HT phase, where the intensity decreases to around 7.6 eV and each feature becomes broadened. Fig. 2(d) shows the spectra obtained by dividing the intensity by LT / HT in Fig. 2(c) assuming a contribution of background features of the HT phase. The analogous asymmetric line shape is found in the Fano profile in resonant photoemission spectroscopy which is given as an interference between the direct photoemission and the Auger processes as a function of the photon energy [16,18]. Fano resonance was first established in a free atom [13] and later observed and applied to surface state [25], adsorbed atoms [19], and molecules [26] on the surfaces. To analyze the Fano profile, the spectrum is fitted by the following Fano formula:

$$I = c_1 \frac{(\varepsilon + q)^2}{1 + \varepsilon^2} + c_2$$

Here, $\varepsilon = 2(E - E_0)/\gamma$, $q$ is the asymmetric parameter, $E_0$ is resonance energy, $\gamma$ is line width, $c_1$ is the amplitude of resonance, and $c_2$ is a background factor. Using Fano-line profile fitting as shown in Fig. 2(d), we obtain $E_0 = 7.57 \pm 0.82$, $q = -1.16 \pm 0.02$, and $\gamma = 0.13 \pm 0.14$. The factor $\gamma$ is related to the lifetime of the excited resonance state, where the value of 5.1 fs is longer than that of photogenerated hole obtained by Lorentizan spectral tail from the high-resolution spectrum of the ML [24].

Hereafter, to address the origin of the observed Fano profile as a function of the photon energy and the related continuum state of PEN/HOPG, the LE-ARUPS E-$k$ map is analyzed. Figs. 3(a)-(d) show the E-$k$ maps where dispersive prominent electronic structures are observed for the LT phase. Corresponding E-$k$ maps for the HT phase are shown in Figs. 3(e)-(h) (for other excitation energies, see Fig. S5). In particular, a convex (hole-like) band with the CFS feature is observed for LT phase, while the flat-discrete band of the HOMO is overlapping at constant binding energy with increasing photon energy. At an excitation energy of 7.4 eV, a convex band with positive contrast (POS) is observed where the maximum energy point appears around 100 meV above the HOMO (00) band at the Γ point. This POS contrast becomes weaker at hν = 7.6 eV and recovers at hν = 7.9 eV. At the excitation energy of 7.6 eV, another band with negative contrast (NEG) appears at the Γ point and becomes more intense at hν = 7.9 eV. Taking into account the Fano profile in Fig. 2(d), the POS and NEG band is observed clearly on each resonance channel depending on the excitation energy. All these dispersive structures are mapped together in the kinetic energy scale as shown in Fig. 2(i), hence a constant final state (CFS) feature appears, indicating the dispersive band contains information about the CB above the $E_v$. The results demonstrate clearly that LE-ARUPS measurement explores the band structure in the region where the secondary electron emission is dominant.

Taking into account the periodic structures observed by LEED in crystalline monolayers (structures A to D), the dispersive band is attempted to reproduce by a simple one-dimensional tight-binding approximation:

$$E(k) = E_c + 2t \cos(ak)$$

where $E_c$ is the position of the band centre, $t$ is the transfer integral, and $a$ is the lattice constant. Parameters of $E_c$ and $t$ are obtained through the least square fitting of the corresponding plot in

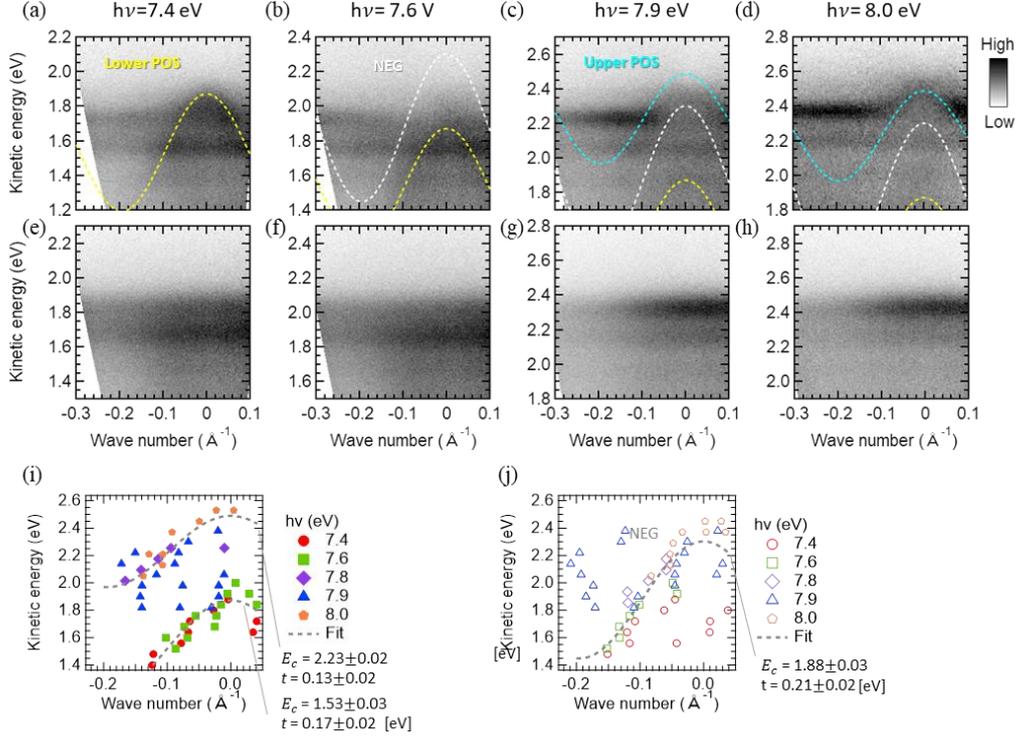

Fig. 3. Photon energy dependence of E-*k* map of PEN/HOPG taken at LT (50 K for 8.0 eV and 14 K for others in top panel) and HT (165 K in middle panel) phase taken at photon energies of 7.4 eV, 7.6 eV, 7.9 eV, and 8.0 eV from left to right, respectively. (bottom) Kinetic-energy position of the dispersive band components with (i) POS and (j) NEG contrast obtained through minima and maxima in second derivatives of the momentum distribution curves of E-*k* maps (h$\nu$=7.4, 7.6, 7.8, 7.9, and 8.0 eV) together with the fitting of a simple one-dimensional tight-binding approximation shown as lines dashed curves.

E-*k* map shown in Figs. 3(i) and (j). Here three dispersive features of lower and upper POSs and one NEG band are analyzed. We used the lattice constant of 16.1 Å. This band structure is neither different from that of other polymorphs of PEN single crystals [27–29], nor that of the K and M point features of graphite in the corresponding energy region (Fig. S4), indicating the dispersive bands are not Umklapp scattering electrons by periodic array on the surface. These facts reveal that PEN adsorption induces a newly formed state in the CB at the interface between the PEN molecules and graphite with highly ordered structures. Note again, there is no new dispersive feature in the VB as shown in the theoretical calculation of the structures A and D (Fig. S2) as well as the LE-ARUPS (HT phase) of liquid-like PEN/HOPG in Figs. 3(e)-(h), and the dilute-PEN adsorbed system on HOPG which also represented almost no DOS in this energy region (Fig. S5(c)). Furthermore, the energy position of the convex band locates in the gap region of the CB of bulk graphite in the previous reports both for the experiment and theory [12,20,21]. To depict the CB above the $E_v$, we have tested preliminary high-level computation and the results indicate some CB features related to experimental convex features both for PEN/graphene and pure graphene (Fig. S6). The quantitative periodicity is dicussed clearly because of faint band features in the experiments as well as the azimuthal rotational randomness of the HOPG.

Here a possible scheme of the photoionization transition of PEN/graphite is shown in Fig. 4. As in Fig. 4(a), the energy conservation matches only for narrow momentum range around Γ point to give a cosine-like CB feature. In other high-momentum regions, FC progressions are observed to transit in a free-electron-like potential. In the case of the liquid-like monolayer at the HT phase, the wavefunction connection is expected in each molecular adsorbate for the CB, though assuming a similar adsorption distance between the HT and LT phases in the vdW approximation. No clear CB features are observed in the HT phase due to a lack of periodicity as depicted in Fig. 4(b). However, the non-trivial energy and momentum dependences of FC factor found in HOMO(00) intensity demonstrate a strong coupling of the molecular wavefunction to graphite. This can be deduced from that the Fanoprofile as a function of photon energy is broader at the HT phase in Fig. 2(c). The origin of the NEG band, which phenomenologically demonstrates the absorption phenomena in the photoelectron emission via Auger-like decay processes [15], is not clearly understood at the present results due to difficulty in the theoretical simulations and the experiments. The characteristics of the second POS band are also difficult with the present data sets (see also Fig. S6). We need to be patient with further challenging experiments of excitation energy dependence and so on.

Finally, the momentum dependence of $q$ of the crystalline phase at each excitation energy is analyzed. The characteristic asymmetric Fano profile is obtained as follows. First, EDC spectra of PEN/HOPG is obtained from the E-$k$ map at different excitation energy (Figs. 5 (a) – (c)). The peak shape around the HOMO (00) is modified significantly at k = 0.00 Å, while the whole spectral shape is modified less at higher wave numbers, especially at hν = 7.9 eV. Second, the intensity difference of LE-ARUPS spectra by subtracting the reference spectra of -0.20 Å$^{-1}$ from that of each momentum is obtained as shown in Figs. 5 (d)−(f). Finally, momentum resolved $q'$ is obtained using Fano profile fitting in three excitation energies.

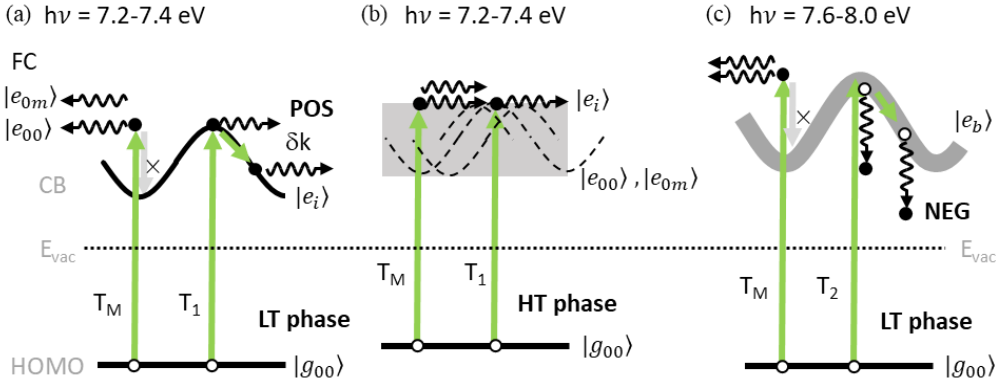

Fig 4. Momentum-resolved transition schemes to describe the ARUPS features. Resonance transitions of (a) between discrete HOMO and the itinerant interface band for the oriented monolayer at the LT phase to give the POS feature and (b) HOMO and the broadened interface band for the liquid-like monolayer at the HT phase observed at lower excitation energies, and (c) resonance of HOMO and the bulk-related band for the oriented monolayer at the LT phase to give the NEG feature observed at higher excitation energies.

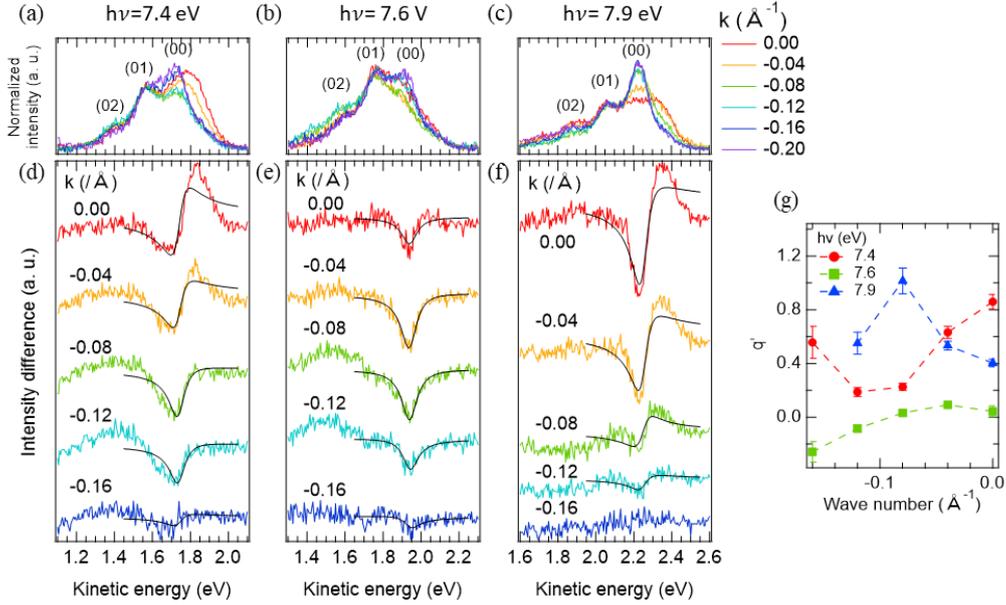

Fig. 5. Photon energy dependence of the LE-ARUPS as from Fig.3. (a–c) Momentum-resolved, background-subtracted spectra of PEN/HOPG taken at photon energies of (a) 7.4 eV, (b) 7.6 eV, and (c) 7.9 eV, respectively, whose intensity is integrated with $k = 0.04$ Å$^{-1}$ and normalized by the intensity at the (01) peak. (d–f) Intensity difference of the momentum resolved spectra by subtracting the spectra of $-0.20$ Å$^{-1}$ from that of other momentums together with the fitting of Fano-line profile for HOMO (00) feature shown as black lines. (g) Asymmetric parameter $q'$ obtained by Fano formula.

The degree of Fano resonance probability can be monitored by the momentum resolved profile $q'$ as a function of wave number for the subtracted EDC spectrum in Fig. 5(g). Unfortunately, we do not discuss the $q'$ parameter for 7.9 eV which may be mixed up the POS and NEG transitions to be complicated. Yet, the other two $q'$ features against the electron momentum are expected either at excitation energies other than or close to the resonance energy. At the excitation energy of 7.4 eV where the POS band feature is observed significantly, the asymmetric parameter $q'$ shows drastic change at $-0.08$ Å$^{-1}$ where the dispersive interface band and the discrete HOMO (00) band is crossing in the E-$k$ map (Fig. 3). On the other hand, at the excitation energy of 7.6 eV where the NEG band is observed significantly close to resonance energy, the almost zero and symmetric profile against the momentum are obtained, where the $q'$ changes less than the case of other excitation energies. The resonance probability to the bulk graphite states, T$_2$ transition in Fig. 2(e) to give the NEG band, is not expected largely and finds almost zero of $q'$. The momentum-resolved Fano profile analysis, which has been done for the first time to our best knowledge, would open a view for the connection of wavefunction, hence being a benchmark of the state-of-the-art theory of quantum simulation.

**Conclusion**

In this study, we have tried to demonstrate how to evaluate the impact of weak interaction on the electronic structure. The oriented monolayer of PEN on graphite (HOPG or single crystal) is an ideal sample for accessing the wave-function overlapping as a typical physisorption system. The ML structure is arranged depending on the temperature from the liquid-like disorder (high-temperature phase) to the incommensurate ML crystal (low-temperature phase). The high-energy and momentum resolutions of LE-ARUPS by scanning the excitation path with synchrotron radiation experiments could demonstrate the impacts of weak interaction on the electronic states. In both ML phases, characteristic asymmetric Fano profile is found against the excitation energy due to the resonant coupling between the discrete molecule-based HOMO state as the initial state and the continuum dispersive band at the interface as the final state. Moreover, in crystalized ML phase, Fano profile analysis has been done for the momentum-resolved LE-ARUPS data. The CFS band originated from a newly-formed conduction band at the physisorbed interface, indicating a strong impact of weak electronic coupling on the connection of the wave function via a larger spread of the unoccupied states demonstrating a fingerprint of measuring the weak interaction at the vdW interface. Asymmetruc parameter $q$ and lifetime at given interference will describe a measurable quantity of weak interaction with further theoretical development in the future.

**ACKNOWLEDGMENTS**

This work was supported in part by the Japan Society for the Promotion of Science (JSPS) KAKENHI (Grant Numbers 26248062, 18H03904, and 20K15176). A part of this work was performed on the BL7U beamline of UVSOR Synchrotron Facility, Institute for Molecular Science (IMS program 15-534, 16-547, 17-547, 18-572, 19-570). We thank staff members of the UVSOR Synchrotron Facility for their support during the measurements. Numerical calculations were performed using the supercomputer facilities at the Cyberscience Center, Tohoku University, and the Institute for Solid State Physics, the University of Tokyo.

**Experiment**

All the experiments were performed at ultra-high vacuum (UHV). A highly oriented pyrolytic graphite (HOPG) and a single crystalline graphite (SCG) were cleaved by a scotch tape before introduction to the UHV chamber. The clean surface was obtained by annealing at 870 K. A molecular layer of PEN was deposited on the substrate at room temperature in a custom-built UHV chamber designed for organic molecular layer deposition. The deposition rate of the PEN was monitored by quartz microbalance and was set to be 0.1– 0.5 Å/min. LEED pattern was measured by a micro-channel plate, a high-sensitive apparatus (OCI). The electronic structure was measured by LE-ARUPS at BL7U in UVSOR Synchrotron Facility, IMS. The excitation energy varied from 7.2 to 8.5 eV, and 21 eV. The total energy resolution was 92 meV, 28 meV, and 10 meV at RT, 165 K, and 14 K, respectively, which was confirmed from the Fermi edge of a reference Au sample. The experiments were carried out separately at several beamtimes, therefore some data at low-temperatures was obtained at different temperatures slightly. However, since each low-temperature data was recorded below the transition temperature, they are identical for Fano resonance analysis and valid for qualitative discussion.

**Calculation**

To clarify the electronic natures of the pentacene-graphite interaction, first-principles electronic-structure calculations were conducted using the program codes [30–33]. The plane waves were used to expand the valence wave function with the energy cutoff of 750 eV, and the projector augmented wave (PAW) potentials were employed to describe the ionic core [34]. The graphene substrates were used with the two-dimensional unit cells of a= b= 16.16 Å, $\gamma$= 120° (structure A) and a= 17.2 Å, b= 6.5 Å, $\gamma$= 79.1° (structure D), which consisted of one and two adsorbate molecules, respectively. The k-point meshes of $2 \times 4$ and $2 \times 2$ were used to sample the two-dimensional Brillouin zone. The periodic slab model with the vacuum layer of 2 nm thickness was used to avoid the artificial interaction along the normal of the slab. The atomic geometries were allowed to relax until all the interatomic forces fell below 0.02 eV/Å. The van der Waals density functional (vdW-DF) was used to describe the intermolecular and molecule-substrate van der Waals forces [35].

**Supporting information**

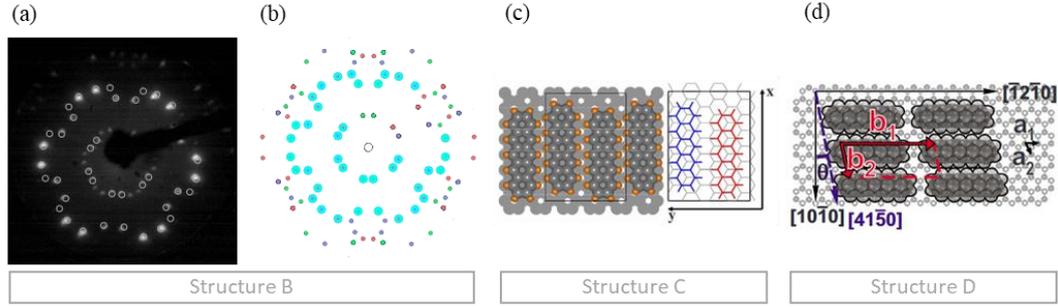

Fig. S1. LEED pattern taken at (a) 72 K with an electron beam energy of 18 eV for another film configurations of PEN/SCG, where the nominal coverage is 4 Å (structure B, see also Fig. S2). The lattice constant is slightly larger than in Fig. 1. Two different structures can coexist. (b) Simulated diffraction pattern depicted considering rotational 3-fold and mirror domains. The lattice parameter was found to be a=18.4 Å, b=6.9 Å and =60.0°. The light blue circles correspond to the diffraction spots observed experimentally. (c) Optimized structure C by theoretical calculation. Ref [5] (Copyright 2008, American Physical Society) (d) Other reported structure D. Ref [21] (Copyright 2011, American Physical Society).

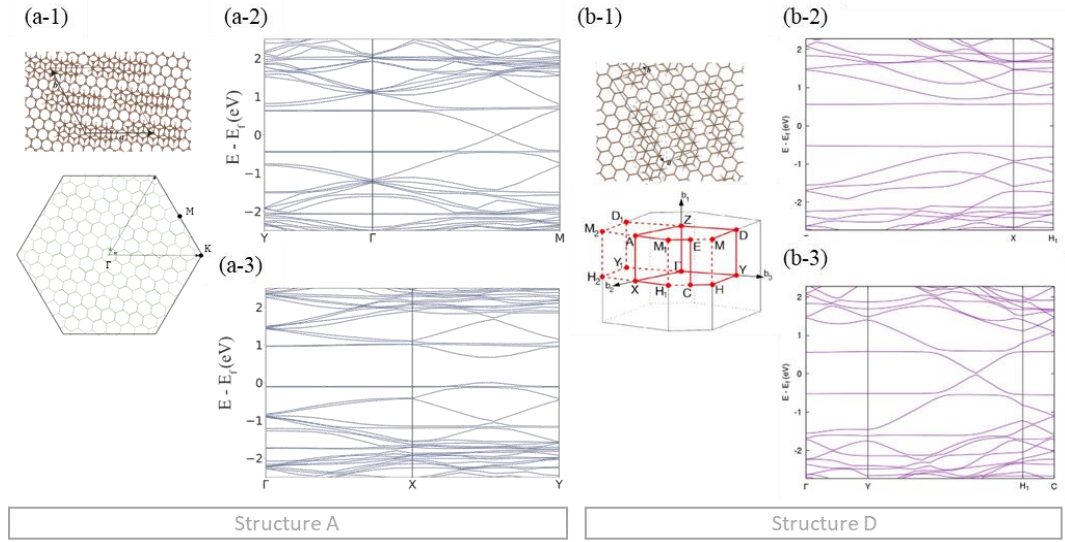

Fig. S2. (a−1) and (b−1): Schematics of the real-space unit cell and the Brillouin zone of structure A and D of the pentacene molecule adsorbed on the graphene substrate, respectively. $\Gamma$-X and $\Gamma$-Y paths correspond to a and b directions in the real-space unit cell, respectively. (a−2) and (a−3): Calculated band dispersions in the pentacene-graphene system along the paths in the Brillouin zone. $\Gamma$(0.0, 0.0, 0.0), X(0.5, 0.0, 0.0), Y(0.0, 0.5, 0.0), and M(0.5, 0.5, 0.0). (b−2) and (b−3): Calculated band dispersions in the pentacene-graphene system along the paths in the Brillouin zone. $\Gamma$ (0.0, 0.0, 0.0), X(0.5, 0.0, 0.0), Y(0.0, 0.5, 0.0), H(0.4815, 0.7407, 0.0), C(0.5, 0.5, 0.0), $H_1$(0.5185, 0.2593, 0.0)

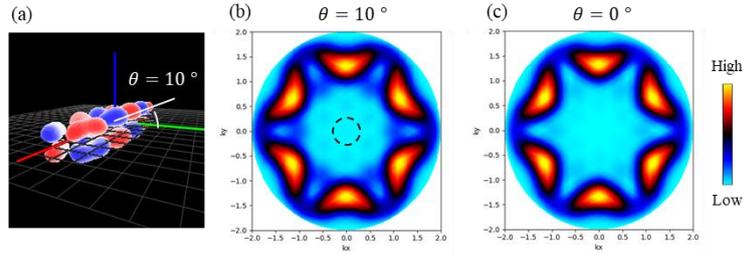

Fig. S3. (a) Adsorption geometry and wave function of molecular orbital of free standing pentacene. Molecular long axis is parallel to the base surface and short axis is inclining 10 ° from the base surface. (b) Simulated momentum map of the HOMO with geometry shown in (a). The simulated momentum maps were obtained by the FT for a free-standing molecule using kMap.py based on a Python program [36]. A three-fold symmetry is applied in the simulation.

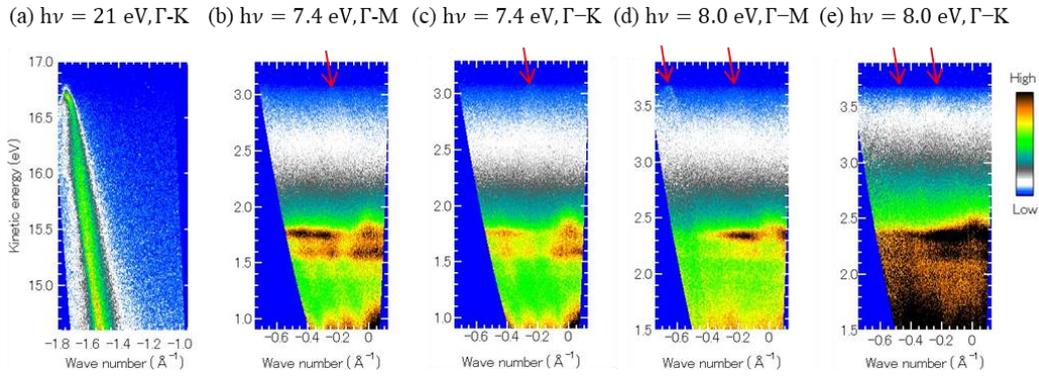

Fig. S4. ARUPS spectra recorded for (a) SCG with h$\nu$ = 21 eV around K point. LE-ARUPS spectra taken for PEN(ML) on the SCG with h$\nu$ = 7.4 eV along (b) ΓM and (c) ΓK, and h$\nu$=8.5 eV along (d) ΓM and (e) ΓK. The surface Umklapp bands are overlapping to the POS and NEG bands.

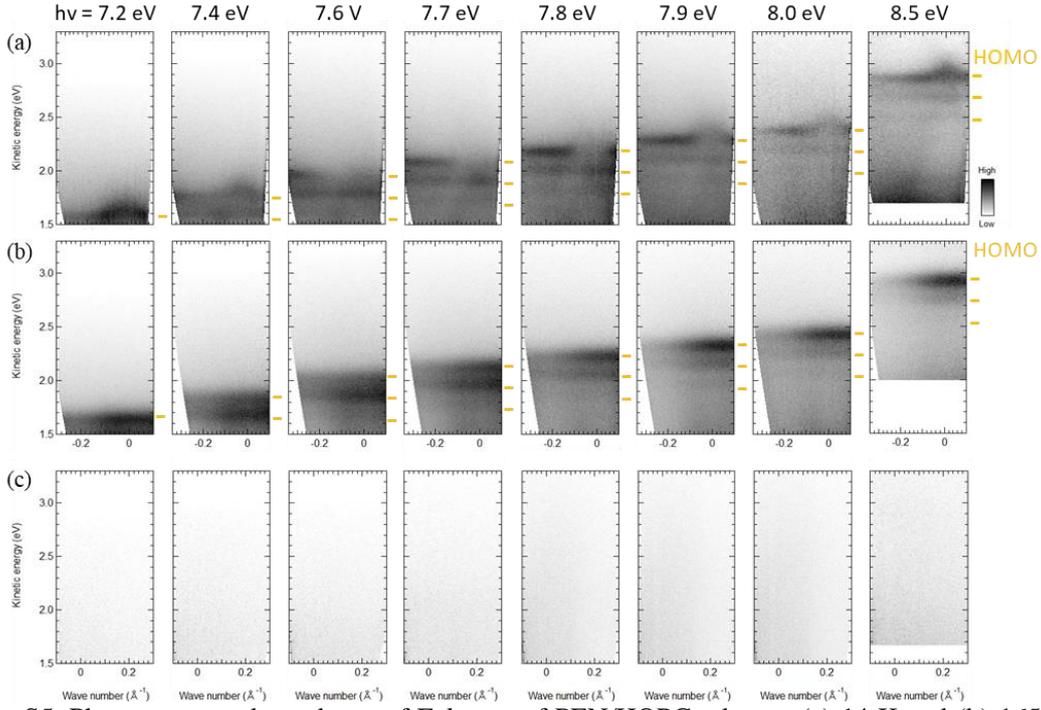

Fig. S5. Photon energy dependence of E-*k* map of PEN/HOPG taken at (a) 14 K and (b) 165 K. Photon energy is varied from 7.2 to 8.5 eV. At 14 K, the convex band with positive contrast is observed around 100 meV above the HOMO (00) band at 7.2 and 7.4 eV especially. With increasing photon energy, the positive band becomes weaker and starts to recover from 7.9 eV. On the other hand, the band with negative contrast is detected in a certain energy region from excitation energy of 7.6 eV to 7.9 eV. (c) Corresponding data are shown for few amount of dilute PEN adsorbates on the HOPG surface, which shows almost no DOS in the VB region.

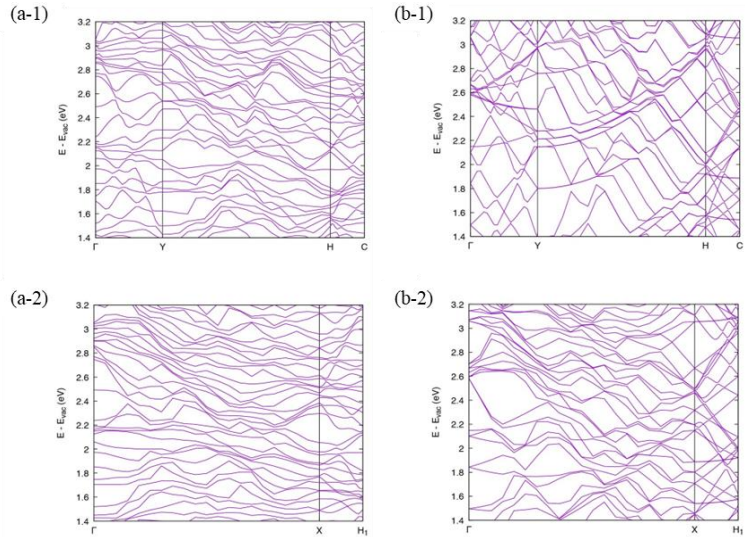

Fig. S6. Calculated band dispersion above the vacuum level of PEN/graphene system of the structure D (a) together with graphene-only system (b). The POS convex bands as found in the experiments is appeaed starting from at 3.0 eV and 2.2 eV along ΓY direction, and 2.9 eV along ΓX direction. We do not discuss quantitatively in detail with the present theoretical level.